\documentclass[prl,twocolumn,reprint,amsmath,amssymb,amsthm]{revtex4-2}
\usepackage{graphicx}
\usepackage{dcolumn}
\usepackage{bm}
\usepackage{bbm}
\usepackage{mathrsfs}
\usepackage{color}
\usepackage{setspace}

\DeclareMathSizes{8}{10}{10}{10}

\definecolor{LinkColor}{RGB}{46,48,146}

\usepackage{hyperref}
\hypersetup{
colorlinks=true,
citecolor=LinkColor,
linkcolor=LinkColor,
urlcolor=LinkColor
}

\begin{document}

\title{Windmill Spin Dynamics and Its Induced Anomalous Hall Effect in Noncollinear Antiferromagnet Mn\textsubscript{3}Sn}

\author{Jikun Zhou}
\affiliation{CAS Key Laboratory of Strongly-Coupled Quantum Matter Physics and Department of Physics, University of Science and Technology of China, Hefei, Anhui 230026, China
}

\author{Yang Gao}
\email[Correspondence author:~~]{ygao87@ustc.edu.cn}
\affiliation{CAS Key Laboratory of Strongly-Coupled Quantum Matter Physics and Department of Physics, University of Science and Technology of China, Hefei, Anhui 230026, China
}
\affiliation{Hefei National Laboratory, University of Science and Technology of China, Hefei 230088, China}

\author{Qian Niu}
\affiliation{CAS Key Laboratory of Strongly-Coupled Quantum Matter Physics and Department of Physics, University of Science and Technology of China, Hefei, Anhui 230026, China
}

\date{\today}

\begin{abstract}
 We demonstrate that the spin dynamics of the noncollinear antiferromagnet Mn\textsubscript{3}Sn hosts a soft eigenmode that transitions from small-angle oscillation to a large-angle, chiral windmill precession once the canting angle exceeds a threshold set by the bending of the spin order. This windmill precession has a fixed chirality of motion, which couples to conduction electrons via a Berry connection polarizability in the mixed space of momentum and spin order, generating a nonzero Berry curvature. This Berry curvature produces a time-independent, DC anomalous Hall effect that persists throughout windmill precession, in sharp contrast to the vanishing Hall response of the static equilibrium spin order when the Hall plane coincides with the Kagome plane. Our results establish a direct link between spin group symmetry, nonlinear spin dynamics, and quantum geometric transport in noncollinear antiferromagnets.
\end{abstract}

\maketitle

Noncollinear antiferromagnets have attracted great interest recently due to their potential applications in spintronics~\cite{AFMRMP2018}. As an outstanding example, Mn\textsubscript{3}Sn possesses a unique chiral spin texture which can couple to quantum geometric properties of Bloch electrons, producing rich magnetotransport phenomena, such as large anomalous Hall effect~\cite{Nakatsuji2015,Kubler2014,Acheche2019,Sung2018,Song2020,Ikeda2018,Lixiaokang2023,Yang2024,Matsuda2020,Singh2020}, spin Hall effect~\cite{Muduli2019,Chen2021}, anomalous Nernst effect~\cite{Guo2017,Lixiaokang2017,Ikhlas2017} and so on. While transport properties due to equilibrium spin texture are well established, the nonequilibrium spin dynamics of  Mn\textsubscript{3}Sn, especially its coupling to quantum geometry, remains much less understood.

Symmetry places strong constraints on the spin dynamics. The spin group of  Mn\textsubscript{3}Sn dictates the magnetic energy to be fully isotropic under a rigid-body in-plane rotation of the spin order up to second order in the spin-orbit coupling~\cite{Liu2026}. Within harmonic approximation, this near-isotropy produces a soft eigenmode, whose gap is set entirely by the intrinsic bending of the spin order~\cite{Dasgupta2020}. This soft-mode physics has so far been treated only in the small-oscillation regime. Large-angle spin dynamics, by contrast, has generally been studied as a driven, dissipative process in the context of switching~\cite{Sun2000, Gomonay2010, Wadley2016, Behovits2023, Polley2023}, and in ${\rm Mn_3Sn}$ specifically it has been demonstrated using external strain and spin torque~\cite{Takeuchi2021, Higo2022, Yoon2023, He2024, Takeuchi2025, Sakamoto2025}. Inherent and switching dynamics therefore remain distinct regimes with different behavior. This raises a fundamental question stemming from the weak anisotropic magnetic energy landscape: can the inherent eigenmode in pristine Mn\textsubscript{3}Sn itself exhibit large-angle dynamics, as seen in switching, and if so, how can such a mode be efficiently excited and detected?

In this work, we answer this question in the affirmative. First, we demonstrate that pristine Mn\textsubscript{3}Sn already possesses large-angle windmill precession as an intrinsic eigenmode of its spin dynamics. By isolating the fast and slow modes in the Landau-Lifshitz equation, the fast bending degree of freedom renormalizes the slow windmill degree of freedom into an effective potential with sixfold rotational symmetry, appearing only at sixth order in spin-orbit coupling. The resulting slow-mode dynamics maps exactly onto a pendulum problem with a low energy barrier. The canting angle serves as canonical momentum and hence governs the behavior of the windmill mode: for small canting, the mode oscillates with a frequency that softens toward zero due to the anharmonic potential; above a threshold canting angle, this oscillation gives way to unidirectional, large-angle windmill precession. Because the threshold is set by a sixth-order anisotropy, it is parametrically small, making the windmill mode readily accessible via a rotating magnetic field.

Crucially, the windmill precession is distinguished from ordinary small-angle oscillation by a fixed chirality of motion. Such dynamically generated chirality can be sensed by electrons, producing a net Berry curvature via a Berry connection polarizability defined in the mixed space of momentum and spin-order angle, in direct analogy to the nonlinear Hall effect. It can manifest as a steady, DC anomalous Hall response during windmill precession, which we refer to as the windmill anomalous Hall effect. In comparison, when the Hall plane coincides with the Kagome plane, the equilibrium Berry curvature and related Hall signal are symmetry-forbidden. Our work reveals the intricate connection between spin group symmetry, spin dynamics, and quantum geometry in noncollinear antiferromagnets, pointing to spin dynamics as a general route to accessing quantum geometric responses beyond those available at equilibrium.

\begin{figure}[t]
	\includegraphics[width=\columnwidth]{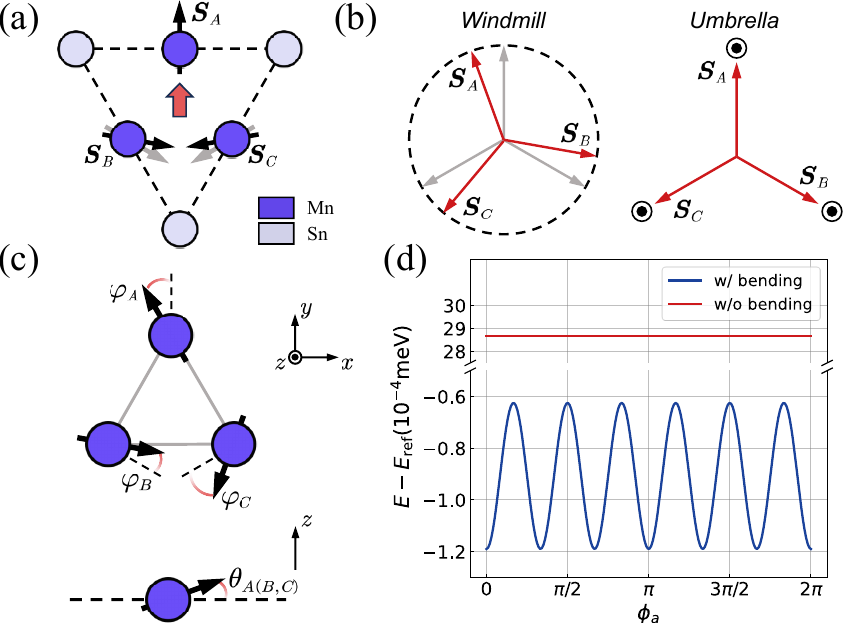}
	\caption{The spin order in ${\rm Mn_3Sn}$. (a) The equilibrium spin order with possible bending from gray to black arrows. (b) Variation of the spin order in the windmill and umbrella modes. (c) Parametrization of spin order variation using spherical angles. (d) Energy variation under rigid-body in-plane rotation of the spin order. The red curve neglects the bending effect, while the blue curve includes it. Here $E_{\mathrm{ref}}=-21.877$ meV is a constant plotting offset introduced to highlight the weak sixfold anisotropy.}
	\label{fig1}
\end{figure}

\textit{Windmill spin dynamics in Mn\textsubscript{3}Sn.}---We first establish the spin Hamiltonian in ${\rm Mn_3Sn}$. Mn\textsubscript{3}Sn crystallizes in a hexagonal lattice with space group $P6_3/mmc$ and point group $D_{6h}$~\cite{MP1197606}. Its structure can be viewed as a layered Kagome lattice with the local spin order in three sublattices making a $120^\circ$ angle between each other within the Kagome plane. By taking into account crystal symmetry and magnetic structure, we use the following two-dimensional spin Hamiltonian~\cite{Liu2026}
\begin{align}
  \label{spinHamiltonian}
    &H _s=J\sum_{\langle \ell \kappa ,\ell ^{\prime}\kappa ^{\prime}\rangle}{\boldsymbol{S}_{\ell \kappa}\cdot \boldsymbol{S}_{\ell ^{\prime}\kappa ^{\prime}}}+J_z\sum_{\langle \ell \kappa ,\ell ^{\prime}\kappa ^{\prime}\rangle}{S_{\ell \kappa, z}S_{\ell ^{\prime}\kappa ^{\prime},z}}\nonumber\\
    &+D\hat{z}\cdot \sum_{\langle \ell \kappa ,\ell ^{\prime}\kappa ^{\prime}\rangle}{\boldsymbol{S}_{\ell \kappa}\times \boldsymbol{S}_{\ell ^{\prime}\kappa ^{\prime}}}+\sum_{\ell \kappa}{\Gamma _{\ell \kappa}^{\alpha \beta}S_{\ell \kappa, \alpha}S_{\ell\kappa,\beta}},
\end{align}
where $\ell,\ell^\prime$ and $\kappa,\kappa^\prime$ are the indices of unit cell and sublattice respectively, and $\alpha,\beta$ denote the spatial direction $x,y,z$. The first term and second term are the isotropic and anisotropic Heisenberg exchange interaction respectively. The third term is the Dzyaloshinskii–Moriya interaction. The last term represents the single-ion anisotropy term, where $\Gamma _{\ell \kappa}$ is a second-order tensor determined by the symmetry of Mn\textsubscript{3}Sn and is of second order in the spin-orbit coupling~\cite{Liu2026}. Currently, there are debates on the ground-state spin configuration of ${\rm Mn_3Sn}$~\cite{Tomiyoshi1982a,Tomiyoshi1986,Duan2015,Xu2026,Zhang2013,Nyari2019}. We use the spin configuration in Fig.~\ref{fig1}(a), which is consistent with the first-principles results~\cite{Liu2026, Zhang2013}. We note that our major results are unaffected by different ground-state spin configurations.

To derive the spin dynamics, we rewrite the Landau-Lifshitz equation in the general Niu-Kleinman form~\cite{Niu1998, Niu1999} using generalized coordinates $\bm{X}_{\ell}$~\cite{suppl}
\begin{equation}
  \label{NKeq}
  \hbar\sum_{\ell^\prime}{\Omega _{\ell \ell ^{\prime}}\dot{\bm X}_{\ell ^{\prime}}}=\frac{\partial H_s}{\partial \bm X_{\ell}},
\end{equation}
where $\Omega_{\ell,\ell^\prime}$ is the Berry curvature between the parameter $\ell$ and $\ell^\prime$ determined by the gyromagnetic ratio. The six parameters are obtained as follows: we parametrize the local spin order by $\varphi_{\ell,\kappa}, \theta_{\ell,\kappa}$ respectively, as shown in Fig.\ref{fig1}(c); by utilizing the irreducible representations of the point group $D_3$, these angles can be re-organized into six normal modes~\cite{Dasgupta2020}, namely the windmill mode $\phi_{a} = (\varphi_{A} + \varphi_{B} + \varphi_{C})/3$, the bending mode $\phi_{b} = \varphi_{C} - \varphi_{B}$, the rocking mode $\phi_{c} = 2\varphi_{ A} - \varphi_{B} - \varphi_{C}$, the umbrella mode $\theta_{a} = (\theta_{A} + \theta_{B} + \theta_{C})/3$, the twisting mode $\theta_{b} = \theta_{ C} - \theta_{B}$, and the wagging mode $\theta_{c} = 2\theta_{A} - \theta_{B} - \theta_{C}$. The windmill mode and the umbrella mode are illustrated in Fig.~\ref{fig1}(b). The parameters in the spin dynamics are then $\bm{X} = (\phi_a, \theta_a, \phi_b, \phi_c, \theta_b, \theta_c)^T$.

These modes can be further grouped by noting that the windmill mode $\phi_a$ is associated with the rigid-body in-plane rotation of the spin order and hence shall have a much smaller energy scale, or equivalently, is very slow dynamically. We therefore separate the variables into slow and fast sectors: $\bm{X}_S=(\phi_a,\theta_a)^T$ and $\bm{X}_F=(\phi_b,\phi_c,\theta_b,\theta_c)^T$. We pair $\theta_a$ with $\phi_a$ since they transform in the same way under symmetry operations. The Niu-Kleinman equations now have the following form
\begin{equation}
  \label{SFNKequation}
  \left( \begin{matrix}
	\Omega _{SS}&		-\Omega _{SF}^{T}\\
	\Omega _{SF}&		\Omega _{FF}\\
\end{matrix} \right) \left( \begin{array}{c}
	\dot{\bm{X}}_S\\
	\dot{\bm{X}}_F\\
\end{array} \right) =\left( \begin{array}{c}
	\partial_{\bm{X}_S} H _s\\
	\partial_{\bm{X}_F} H _s\\
\end{array} \right)
\end{equation}
where $\Omega_{SS}$, $\Omega_{FF}$ and $\Omega_{SF}$ are spin Berry curvatures within the slow and fast degrees of freedom and cross slow and fast modes. 

The fast modes can be treated in the harmonic approximation. Therefore, the magnetic energy shall take the following form~\cite{suppl}:
\begin{equation} \label{eq_eng}
  H _s=\bm{X}_{F}^{T}f\left( \bm{X}_S \right) \bm{X}_F+g^T\left( \bm{X}_S \right) \bm{X}_F+h\left( \bm{X}_S \right),
\end{equation}
where $f(\bm{X}_S)$ is a $4\times 4$ symmetric matrix, $g(\bm{X}_S)$ is a $4\times 1$ vector, and $h(\bm{X}_S)$ consists of a constant term and a leading quadratic term in $\theta_a$, while being independent of $\phi_a$ as shown in Fig.~\ref{fig1}(d)~(red). Hence, the magnetic energy is also quadratic with respect to $\theta_a$ since its energy scale is on the same order with $\bm X_F$. Such form is nevertheless incorporated in Eq.~\eqref{eq_eng}.  

The bending effect originates from the linear term in the magnetic energy. Using the condition $\partial H_s/\partial \bm{X}_F = 0$ we obtain
\begin{align}
    \bm{X}_{F}^{\mathrm{eq}} = -\frac{1}{2}f^{-1}\left( \bm{X}_S \right) g\left( \bm{X}_S \right)\,.
\end{align}
Only the bending component of $\bm {X}_F^{\mathrm{eq}}$ is nonzero and since $f$ and $g$ vary with the rigid-body rotation of the spin order, so does the bending. The bending effect offers the potential energy for the dynamics of $\phi_a$. To see this, we define the effective variation of the fast mode as $\tilde{\bm{X}}_{F}=\bm{X}_F - \bm{X}_{F}^{\mathrm{eq}}$. The energy for $\tilde{\bm{X}}_{F}$ does not contain linear terms so that the dynamics of $\tilde{\bm{X}}_{F}$ is purely harmonic. Accordingly, the constant energy changes to~\cite{suppl}
\begin{align}
    h(\bm{X}_S)\rightarrow \tilde{h}(\bm{X}_S)=h(\bm{X}_S)-\frac{1}{4} g^T(\bm{X}_S) f^{-1}(\bm{X}_S) g(\bm{X}_S)\,.
\end{align}
By taking into account the bending effect $\tilde{h}(\bm{X}_S)$ is now anisotropic as shown in Fig.~\ref{fig1}(d).

The fast and slow modes can then be solved numerically. The inter-mode coupling can in fact be ignored so that the fast and slow modes can be isolated. The reason is that the inter-mode dynamical matrix $\Omega_{SF}$ couples two degrees of freedom that transform differently under point group operations. Therefore, it depends on $\tilde{\bm{X}}_F$ to make the equations of motion covariant, and is hence at least one order of magnitude smaller. By solving the $4$-by-$4$ equations of motion for the fast mode, we obtain the eigen frequencies $\nu_{F,1}=2.70$ THz and $\nu_{F,2}=2.81$ THz which agree well with the numerical result~\cite{suppl}, confirming the validity of our analysis. 

\begin{figure}[t]
	\includegraphics[width=\columnwidth]{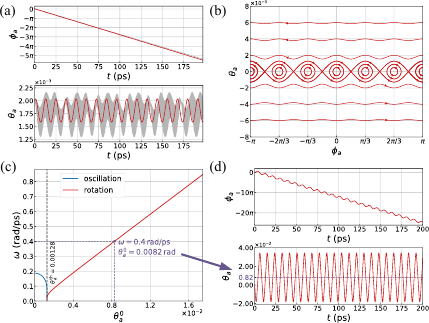}
	\caption{Eigen modes of spin dynamics in ${\rm Mn_3Sn}$. In (a), the numerical and analytical results are shown by gray and colored lines, respectively. (b) Phase diagram of $\phi_a$ and $\theta_a$ with arrows denoting the direction of motion. (c) Eigen frequency with different initial values of $\theta_a$. (d) Excitation of an eigen mode using an in-plane microwave with $\bm{B}(t)=B_0(\sin\omega_B t, \cos\omega_B t, 0)$~($\omega_B = 0.4$ rad/ps and $B_0=86.4$ mT). The purple dashed line shows the mean value of $\theta_a$.}
	\label{fig2}
\end{figure}

The numerical solution of the slow mode is shown in Fig.~\ref{fig2}(a). The windmill precession can be directly seen from the ever decreasing behavior of $\phi_a$, with a period consistent with the average value of the canting angle. The umbrella mode consists of a high-frequency oscillation superimposed on a low-frequency component over a long time window. The frequency of the latter is six times of that of $\phi_a$.

To further understand the feature of the windmill precession, we start from the isolated equations of motion $\Omega_{SS}\dot{\bm{X}}_S=\partial\tilde{h}/\partial \bm{X}_S$. We calculate $\partial\tilde{h}/\partial \bm{X}_S$ numerically and up to leading order it has the form~\cite{suppl}: $\partial\tilde{h}/\partial \bm{X}_S=(a_0 \sin (6\phi_a), a_1 \theta_a)^T$, where fitting parameters $a_0\propto \Gamma_{12}^3$ and $a_1$ mainly depends on $J$, $J_z$ and $D$. Since $\Gamma_{12}$ is at second order of the spin-orbit coupling, the anisotropic potential energy is then at sixth order and hence quite small. The equations of motion of slow modes then have the form~\cite{suppl}
\begin{equation}
  \left( \begin{array}{c}
	\dot{\phi}_a\\
	\dot{\theta}_a\\
\end{array} \right) =\frac{1}{3}\left( \begin{array}{c}
	-a_1\theta _a\\
	a_0\sin \left( 6\phi _a \right)\\
\end{array} \right).
\end{equation}
This dynamics can be reformulated as a Hamiltonian system, by treating $\theta_a$ and $\phi_a$ as canonical momentum and position respectively, with the energy given by
\begin{equation}
  \label{Heff}
H _{\mathrm{eff}}=\frac{1}{3}\left[ \frac{1}{2}a_1{\theta _a}^2-\frac{a_0}{6}\cos \left( 6\phi _a \right) \right],
\end{equation}

This effective Hamiltonian describes a pendulum-like system with six-fold periodic potential. Its dynamical phase diagram is shown in Fig.~\ref{fig2}(b), which has two distinct types of behaviors. When the total energy is negative, the magnetic order will oscillate around the stable point and the period can be calculated by elliptic integral~\cite{Sala1989}, which gives $T \approx 33.6$ ps~\cite{suppl}. As the canting angle increases, the period quickly diverges as shown in Fig.~\ref{fig2}(c). There is a threshold value for the canting angle determined by the bending effect: 
\begin{equation}
    \theta_a^{th}=\sqrt{2|a_0|/3a_1}\approx 0.00128\,.
\end{equation}
$\theta_a$ is at the third order of the spin-orbit coupling.
Above this value, the kinetic energy can surpass the potential barrier, so that $\phi_a$ varies uni-directionally, producing a windmill precession~\cite{suppl}. For relatively large canting angle, $\dot{\theta}_a$ can be ignored, and $\dot{\phi}_a$ is then fully determined by the canting angle, leading to a linear dependence of the eigen frequency on the canting angle as shown in Fig.~\ref{fig2}(c), which is also consistent with numerical result in Fig.~\ref{fig2}(a). The windmill precession can be easily excited using a rotating magnetic field as shown in Fig.~\ref{fig2}(d).


\textit{Windmill anomalous Hall effect.}---The windmill mode is a chiral motion with large magnitude. Therefore, it can have a significant impact on electronic transport. 

According to the Onsager's reciprocal relation, the anomalous Hall effect requires a chiral direction to break the time-reversal symmetry. Usually, the chiral direction is provided by the equilibrium spin order. However, in ${\rm Mn_3Sn}$, when the Hall plane lies in the Kagome plane, there is a combined $\mathcal{TM}_z$ symmetry, which forbids the anomalous Hall effect.

When the windmill mode is excited, its chiral direction can be sensed by electrons and hence leads to a nonzero anomalous Hall effect in the Kagome plane. To show this, we start from the formula of the anomalous Hall conductivity vector~\cite{NagaosaRMP2010}:
\begin{align}
    \sigma^{AH}_{\alpha}&=-\frac{e^2}{\hbar} \sum_n\int \frac{d^2\bm k}{(2\pi)^2} f_{n\bm{k}}\Omega^{\alpha}_{n\bm{k}} \,,
\end{align}
where $f_{n\bm{k}}$ is the Fermi distribution function, and $\Omega^{\alpha}_{n\bm{k}}$ is the Berry curvature in the momentum space~\cite{XiaoRMP2010}. When the spin order is rotated in accordance with the windmill mode, the electronic wave function is time-dependent. Within the adiabatic assumption, for the wave function in the $n$-th band we have $|\psi_n\rangle=|\psi_n(\phi_a(t))\rangle$. This makes the Berry curvature time-dependent. However, its net effect over a cycle vanishes identically as dictated by the $\mathcal{TM}_z$ symmetry.

To obtain the driven effect of the chiral mode, we plug the instantaneous wave function in the Schrodinger equation, and obtain
\begin{align}
    i\hbar\partial _t|\psi \rangle =\left( \hat{H}_e-i\hbar\dot{\phi}_a\partial _{\phi_a} \right) |\psi \rangle\,,
\end{align}
where $\hat{H}_e$ is the electronic Hamiltonian. Using the perturbation theory, we can then solve $|\psi_{n\bm{k}}\rangle$ up to the first order in the angular velocity $\dot{\phi}_a$. The result is $|\tilde{\psi}_{n\bm{k}}\rangle=|\psi_{n\bm{k}}\rangle+|\delta \psi_{n\bm{k}}\rangle$, where
\begin{align}
    |\delta \psi_{n\bm{k}}\rangle=e^{i\bm k\cdot \bm r} \sum_{m\neq n}|u_{m\bm{k}}\rangle \frac{\hbar\dot{\phi}_a \langle u_{m\bm{k}}|i\partial_{\phi_a}|u_{n\bm{k}}\rangle}{\varepsilon_{m\bm{k}} -\varepsilon_{n\bm{k}}}\,,
\end{align}
$|u_{n\bm{k}}\rangle$ is the periodic part of the instantaneous Bloch state and $\varepsilon_{n\bm{k}}$ is the instantaneous band energy. Using the perturbed wave function, we can calculate the correction to the Berry connection $\bm{\mathcal{A}}_{n\bm{k}}= \bm G^n_{\bm{k}\phi_a}\hbar\dot{\phi}_a$, where
\begin{align}
     \label{eq_bcp} \bm G^n_{\bm{k}\phi_a}&=2\mathrm{Re}\sum_{m \neq n}{\frac{\langle u_{n\boldsymbol{k}}|i\partial _{\boldsymbol{k}}|u_{m\boldsymbol{k}}\rangle \langle u_{m\boldsymbol{k}}|i\partial _{\phi_a}|u_{n\boldsymbol{k}}\rangle}{\varepsilon _{m\boldsymbol{k}}-\varepsilon _{n\boldsymbol{k}}}}\,.
\end{align}
In analogy with the nonlinear Hall effect, $\bm G^n_{\bm{k}\phi_a}$ has the meaning of the Berry connection polarizability~\cite{Gao2014,Gao2015,Liu2022}. It is, however, not in the momentum space but in the orientation space of the spin order. The curl of $\bm{\mathcal{A}}_{n\bm k}$ then yields the Berry curvature and hence the anomalous Hall conductivity.

There is an additional contribution to the Hall signal due to the Thouless pumping effect~\cite{Thouless1983}. By taking into account the windmill dynamics, the system is time-dependent, allowing a pumping current given by
\begin{align}
\bm J=e\int \frac{d^2\bm k}{(2\pi)^2} f_{n\bm k}\bm \Omega_{\bm k \phi_a} \dot{\phi}_a\,,
\end{align}
where $\bm \Omega_{\bm k \phi_a}=\partial_{\bm k}\mathcal{A}_{\phi_a}-\partial_{\phi_a} \bm{\mathcal{A}}_{\bm k}$ is the mixed Berry curvature~\cite{Sundaram1999}. Under external electric field, both $\bm{\mathcal{A}}_{\bm k}$ and $\mathcal{A}_{\phi_a}$ changes. However, only the latter can yield a Hall-type signal~\cite{suppl} and it can be expressed using the same Berry connection polarizability: $\delta \mathcal{A}_{\phi_a}= -e\bm G_{\bm k\phi_a}^n \cdot \bm E$~\cite{Chen2024, Li2024}. Using $\delta \mathcal{A}_{\phi_a}$ we then obtain the Hall-type pumping current.
 
\begin{figure}[t]
	\includegraphics[width=\columnwidth]{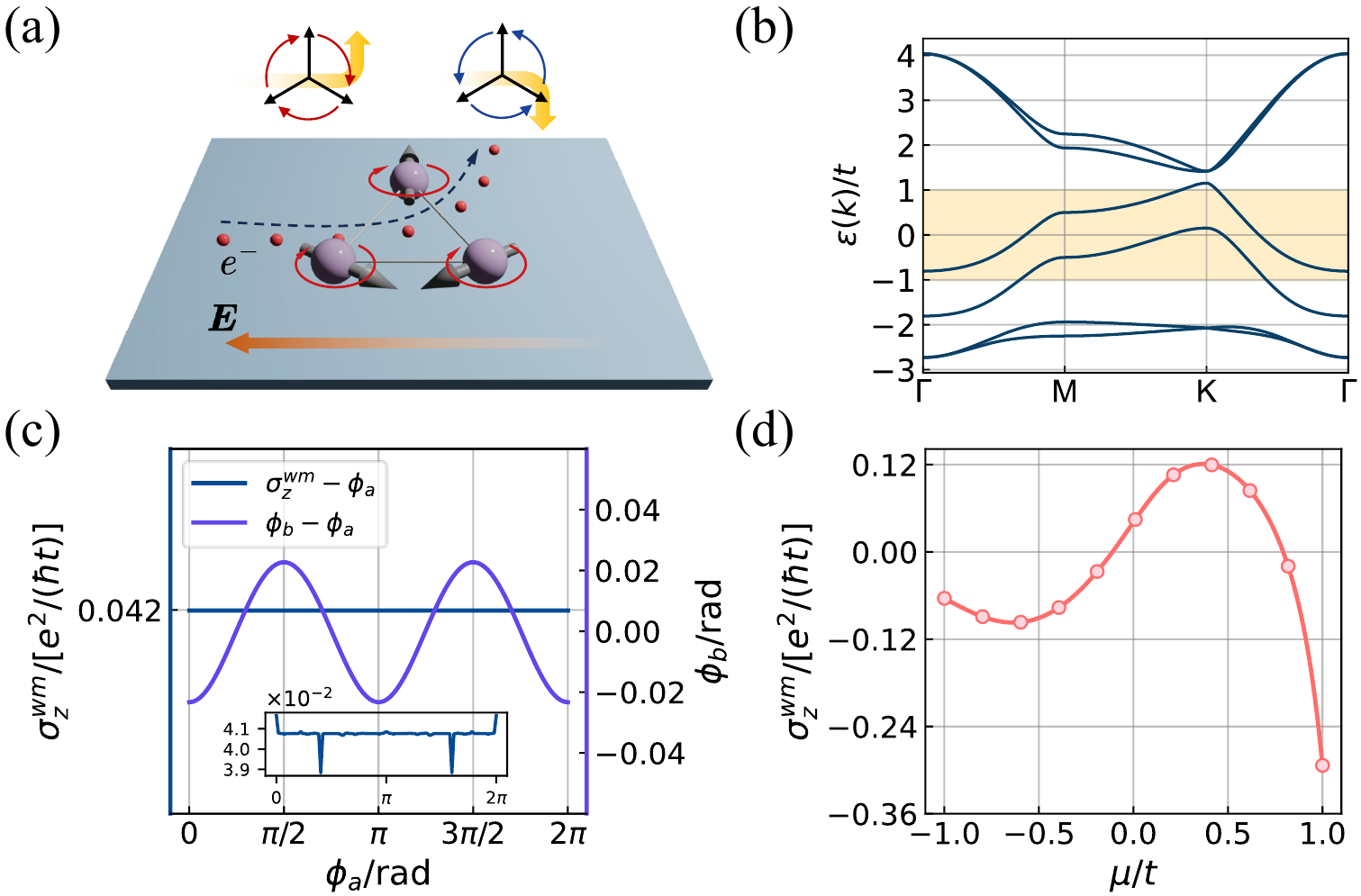}
	\caption{The windmill anomalous Hall effect. (a) The schematic illustration. (b) The electronic band structure corresponding to Hamiltonian in Eq.~\eqref{He}. (c) Dependence of windmill conductivity $\sigma_{z}^{wm}$ and bending angle $\phi_b$ on the windmill angle $\phi_a$. The inset shows the dependence of $\sigma^{wm}_z$ on $\phi_a$ when the variation of $\phi_b$ is taken into account. (d) Dependence of $\sigma^{wm}_z$ on the chemical potential $\mu$. The corresponding range of $\mu$ is highlighted by the yellow shaded region in (b).}
	\label{fig3}
\end{figure}

By combining both contributions, we obtain the final result for the anomalous Hall conductivity driven by the windmill mode~\cite{suppl}: $\sigma_\alpha^{AH}=\sigma_\alpha^{wm} (\hbar\dot{\phi}_a) $ with
\begin{align}
    \sigma_{\alpha}^{wm}=\frac{3e^2}{2\hbar}\sum_n{\int{\frac{d^2\boldsymbol{k}}{\left( 2\pi \right) ^2}\frac{\partial f_{n\boldsymbol{k}}}{\partial \varepsilon _{n\boldsymbol{k}}}\left( \bm{v}_{n\boldsymbol{k}}\times \bm G_{\boldsymbol{k}\phi_a} \right)_\alpha}}\,.
\end{align}
We refer to such effect as the windmill anomalous Hall effect. It flips sign when the chiral direction of the windmill mode is reversed, as illustrated in Fig.~\ref{fig3}(a).

This windmill anomalous Hall effect is not restricted by the $\mathcal{TM}_z$ symmetry. Under the time reversal operation, $\bm k\rightarrow -\bm k$ and $\phi_a\rightarrow \pi+\phi_a$. Therefore, we have $\bm G^n_{\bm{k}\phi_a}(\phi_a)\rightarrow -\bm G^n_{\bm{k}\phi_a}(\phi_a+\pi)$. Since the velocity also flips sign, the windmill anomalous Hall conductivity then transforms as $\sigma_{z}^{wm}(\phi_a)\rightarrow \sigma_{z}^{wm}(\pi+\phi_a)$. Similarly, under the mirror-$z$ operation, $\phi_a\rightarrow \phi_a+\pi$ and the in-plane momentum stays the same. Therefore, $\sigma_{z}^{wm}(\phi_a)\rightarrow \sigma_{z}^{wm}(\pi+\phi_a)$. Therefore, the combined $\mathcal{TM}_z$ symmetry has no restriction on $\sigma_z^{wm}$. We note that the $x$ and $y$ components of $\bm \sigma^{wm}$ are forbidden by the $\mathcal{TM}_z$ symmetry. 

To demonstrate the windmill anomalous Hall effect, it is sufficient to consider a two-dimensional lattice model in the Kagome plane. The Hamiltonian reads as
  \begin{align}\label{He}
    H_e=&t\sum_{\langle \ell \kappa ,\ell ^{\prime}\kappa ^{\prime}\rangle ,s}{c_{\ell \kappa s}^{\dagger}c_{\ell ^{\prime}\kappa ^{\prime}s}}+J_{\mathrm{sd}}\sum_{\ell \kappa ,s,s^{\prime}}{\left( \boldsymbol{S}_{\ell\kappa}\cdot \boldsymbol{\sigma } \right) _{ss^{\prime}}c_{\ell \kappa s}^{\dagger}c_{\ell\kappa s^{\prime}}}\notag\\
    &+i\lambda \sum_{\langle \ell \kappa ,\ell ^{\prime}\kappa ^{\prime}\rangle ,s,s^{\prime}}{\nu _{\kappa \kappa ^{\prime}}\left( \sigma _z \right) _{ss^{\prime}}c_{\ell \kappa s}^{\dagger}c_{\ell ^{\prime}\kappa ^{\prime}s^{\prime}}},
  \end{align}
where $\bm \sigma$ is the Pauli matrix for spin. 
The three terms are the nearest neighbour hopping, the \textit{s-d} exchange coupling, and the intrinsic spin-orbit coupling, respectively. We have chosen $\nu_{AB} = \nu_{BC} = \nu_{CA} = 1$ and $\nu_{\kappa\kappa^\prime} = -\nu_{\kappa^\prime\kappa}$. The band spectrum is shown in Fig.~\ref{fig3}(b).

We calculate the windmill anomalous Hall effect in this model. During the windmill precession of the spin order, $\sigma_{z}^{wm}$ is nearly a constant, as shown in Fig.~\ref{fig3}(c). This clearly demonstrates the DC nature of the windmill anomalous Hall effect. We also refine our calculation, by taking into account the variation of the bending of the spin order during the windmill precession given in Fig.~\ref{fig3}(c). The resulting response coefficient $\sigma_{z}^{wm}$ is nearly steady but with a slightly smaller magnitude, and with small dips at certain windmill angle. 

The windmill anomalous Hall effect also depends sensitively on the chemical potential. Since the Berry connection polarizability depends inversely on the band gap, it is most important near the band extrema where the band gap is small. Moreover, the Berry connection polarizability has opposite signs across a band gap according to Eq.~\eqref{eq_bcp}. All these features are consistent with the pattern of $\sigma_z^{wm}$ in Fig.~\ref{fig3}(d).

In summary, we show the existence of the windmill precession in the inherent spin dynamics of Mn\textsubscript{3}Sn. It emerges as the canting angle is higher than certain threshold value determined by the bending effect and has a frequency linearly dependent on the canting angle. Its unique chiral precession direction can be sensed by electrons to produce a net Berry curvature through the Berry connection polarizability. Such induced Berry curvature can then yield a steady anomalous Hall effect.

\begin{acknowledgments}
    We acknowledge useful discussions with Dazhi Hou, Jiahao Han, Cong Xiao, Zheng Liu and Xiaoqiang Liu. This work is supported by the National Natural Science Foundation of China (12234017). Y. G. is also supported by the Innovation Program for Quantum Science and Technology (2021ZD0302802). The supercomputing service of USTC is gratefully acknowledged.
\end{acknowledgments}

\textit{Data availability.}---The data that support the findings in this manuscript are available from the authors upon reasonable request.

%

\end{document}